\documentstyle[12pt]{ioplppt}
\begin{document}
\title{Kinetic theory for nongeodesic particle motion: 
Selfinteracting equilibrium states and effective viscous fluid pressures
}

\author{Winfried Zimdahl\dag and Alexander B. Balakin\dag\ddag}

\address{\dag\ Fakult\"at f\"ur Physik, Universit\"at Konstanz, 
PF 5560 M678
D-78457  Konstanz, Germany
}

\address{\ddag\ Department of General Relativity and Gravitation\\ 
Kazan State University, 420008 Kazan, Russia
}

\begin{abstract}
The particles of a classical relativistic gas are supposed to move under the influence of a quasilinear (in the particle four-momenta), self-interacting force inbetween elastic, binary collisions. 
This force which is completely fixed by the equilibrium conditions of the gas, gives rise to an effective viscous pressure on the fluid phenomenological 
level. 
Earlier results concerning the possibility of accelerated expansion of the universe due to cosmological particle production are reinterpreted. 
A phenomenon such as power law inflation may be traced back to 
specific self-interacting forces keeping the particles of a gas universe 
in states of generalized equilibrium.  
\end{abstract}


\section{Introduction}
In its most elementary version the kinetic theory of a simple relativistic gas relies on the concept of N pointlike particles which may interact through elastic, binary collisions. 
Inbetween the collisions which are assumed to establish an 
(approximate) local or global equilibrium of the system the particles move on geodesics of either a given spacetime or a spacetime selfconsistently determined by the gas particles themselves. 
Technically, the gas particles are described by an invariant one-particle distribution function governed by Boltzmann's equation. 
The macroscopic fluid dynamics for such a system may be obtained in terms of the first and second moments of the (close-to-equilibrium) distribution function. 
An interesting area of application is the dynamics of a 
cosmological fluid.  
While hydrodynamic concepts such as energy density and pressure are assumed to be useful back to times of the order of the Planck time it is less clear whether the hot and dense early universe  is accessible to a kinetic description as well. 
A gas, however, is the only system for which the correspondence between microscopic variables, governed by a distribution function, and phenomenological fluid quantities is sufficiently well understood. 
This property makes gas universes interesting toy models. 
There exists a considerable body of literature refering to model universes for which a kinetic approach is assumed to be applicable 
(see, e.g.,\cite{Bern,TZP,ZTP} and references therein). 
One hopes that such kind of approach is also able to 
give an idea of the relevant physics in our real Universe. 

The behaviour of structureless particles moving on geodesics inbetween elastic,  binary collisions, the latter being subject to the hypothesis of 
``molecular chaos'', equivalent to neglegting interparticle correlations, is by now well understood. 
Sophisticated solution techniques have been developed and applied to numerous physically relevant situations \cite{Sas,Stew,Ehl,IS,Groot,MaWo}. 
The usual procedure here is first to characterize equilibrium states, i.e., states with vanishing entropy production and to relate the parameters of the corresponding distribution function to macroscopic (perfect) fluid quantities. 
Nonequilibrium situations are then, in a second step, taken into account as deviations from equilibrium. 

The concept of a simple relativistic gas has been refined in various ways. 
For example, the gas particles may be supposed to have internal structure \cite{Isr,FeldTau}. 
Taking into account interparticle correlations in the simplest possible manner is equivalent to an additional stochastic term in Boltzmann's equation, leading to fluctuating hydrodynamics \cite{Zfluc,Zfluc2}. 
In order to investigate cosmological particle production processes, supposedly of quantum origin, ``source'' terms have been introduced into 
Boltzmann's equation giving rise to corresponding sources in the macroscopic balance equations for the extensive thermodynamic quantities 
(particle number, energy, entropy) \cite{TZP,ZTP,WZ}. 
Under certain conditions these macroscopic sources are equivalent to effective viscous fluid pressures. \\
Any of these modifications and extensions of the kinetic theory of a simple gas is supposed to enlarge its ranges of applicability and credibility since it drops at least one of its simplifying assumptions. 

In the present paper we investigate a system of point particles interacting through elastic, binary collisions but not necessarily moving on geodesics inbetween the collisions. 
Geodesic motion of particles is a highly idealized case. 
In reality, particle worldlines are supposed to deviate from geodesics 
since the particles will be subject to additional interactions in general. 
We assume here that these interactions may be modelled as effective forces on the particles. 
The kinetic theory for particles under the influence of various forces was considered, e.g.,  in \cite{Ehl,Groot,Hakim,BaGo}. 
We will focus on equilibrium states of the gas in a 
simple force field. 
It is well known that a Lorentz force which influences the particle motion inbetween the collisions  will change, in general, the conditions under which the system of particles is in equilibrium, but not the functional structure of the equilibrium distribution function. 
Here we are interested in forces which are quasilinear (in a sense described below) in the particle four-momenta. 
We will 
find the most general quasilinear force  compatible with the equilibrium distribution function of a classical Maxwell-Boltzmann gas. 
The spacetime structure of this force 
is not assumed to be given but will be determined from the 
(generalized) equilibrium conditions for the gas. 
One part of the force will turn out to be proportional to the fluid expansion of the system, another one proportional to 
the spatial gradient of $\mu /T$ where $\mu $ is the chemical potential and 
$T$ is 
the fluid temperature. 
It is characteristic for our approach that the force depends both on the microscopic particle momenta and on macroscopic fluid quantities which characterize the $N$-particle system as a whole. 
Consequently, it describes a selfinteraction of the gas. 
Physically, the concept of a selfinteracting force may be regarded as a mapping of parts of the interactions in a many-particle system onto an effective one-particle quantity. 
We expect this concept to be useful in circumventing some of the general problems inherent in attempts to formulate a relativistic statistics for interacting many-particle systems (see, e.g., \cite{IsKan} and references therein). 
The selfinteracting force  
neither preserves the particle number nor the energy momentum and it will give rise to entropy production. 
The corresponding source term in the energy momentum balance may consistently be mapped on the effective viscous pressure of a conserved, imperfect fluid energy momentum tensor. 
This deepens our understanding of the phenomenological description of microscopic processes in terms of effective bulk pressures. 
We will establish the correspondence to previous results \cite{TZP,ZTP,WZ} 
on kinetic theory and particle production which are recovered here in a physically more transparent way and with less assumptions. 
We show that the effective viscous pressure which may considerably modify the dynamics of a  
Friedmann-Lema\^{i}tre-Robertson-Walker 
(FLRW) universe, including the possibility of power law inflation, is related to specific selfinteracting forces on the microscopic gas particles. 
Moreover, for interesting limiting cases the newly introduced force concept allows us to integrate the microscopic equations of motion explicitly and thus to establish an exactly solvable model of a selfinteracting gas universe.

The plan of the paper is as follows. 
In section 2 we establish the general kinetic theory of a relativistic gas under the influence of an arbitrary four-force and recall the concept of ``generalized equilibrium''. 
The corresponding equilibrium conditions for quasilinear forces are obtained in section 3. 
Section 4 is devoted to the fluid dynamics of a selfinteracting gas in generalized equilibrium. 
Earlier results on the role of cosmological particle production in a phenomenological context are reinterpreted and effective viscous pressures 
are related to forces on the microscopic gas particles. 
Section 5 gives a brief summary of the paper. 
Units have been chosen so that $c = k_{B} =  \hbar =1$.

\section{General kinetic theory}

The one-particle distribution function 
$ f = f\left(x,p\right)$ 
for  relativistic gas particles under the influence of a four-force 
$F ^{i}=F ^{i}\left(x,p \right)$ obeys the Boltzmann equation 
\cite{Stew,Ehl,IS}
\begin{equation} 
p^{i}f,_{i} - \Gamma^{i}_{kl}p^{k}p^{l}
\frac{\partial f}{\partial
p^{i}} + m F ^{i}\frac{\partial{f}}{\partial{p ^{i}}}
 = C\left[f\right]   \mbox{ , } \label{1}
\end{equation}
where $f\left(x, p\right) p^{k}n_{k}\mbox{d}\Sigma dP$ 
is the number of
particles whose world lines intersect the hypersurface element 
$n_{k}d\Sigma$ around $x$, having four-momenta in the range 
$\left(p, p + \mbox{d}p\right)$;  $i$, $k$, $l$ ... = $0$, 
$1$, $2$, $3$. Here 
$\mbox{d}P = A(p)\delta \left(p^{i}p_{i} + m^{2}\right) 
\mbox{d}P_{4}$
is the volume element on the mass shell 
$p^{i}p_{i} = -  m^{2} $
in the momentum space. 
$A(p) = 2$, if $p^{i}$ is future directed and 
$A(p) = 0$ otherwise; 
$\mbox{d}P_{4} = \sqrt{-g}\mbox{d}p^{0}\mbox{d}p^{1}
\mbox{d}p^{2}\mbox{d}p^{3}$. The quantity 
$C[f]$ is Boltzmann's collision term. 
Its specific structure discussed e.g. by 
Ehlers \cite{Ehl} will not 
be relevant for our considerations. 
Following Israel and Stewart \cite{IS} we shall only 
require that (i) $C$
is a local function of the distribution function, i.e., 
independent
of derivatives of $f$, (ii) $C$ is consistent with 
conservation of
four-momentum and number of particles, and (iii) $C$
yields a nonnegative expression for the entropy 
production and does
not vanish unless $f$ has the form of a local equilibrium
distribution (see eq.(\ref{7}) below). 
Equation (\ref{1}) implies that the mass-$m$ particles inbetween the 
collisions move according to the equations of motion
\begin{equation}
\frac{\mbox{d} x ^{i}}{\mbox{d} \gamma  } = p ^{i}\ ,
\ \ \ \ \ \ 
\frac{\mbox{D} p ^{i}}{\mbox{d} \gamma  } = m F ^{i}\ ,
\label{2}
\end{equation}
where $\gamma  $ is a parameter along the particle worldlines which 
for $m > 0$ 
may be related to the proper time $\tau $ by $\gamma   = \tau /m$. 
Since the particle four-momenta are normalized according to 
$p ^{i}p _{i} = - m ^{2}$, the force $F ^{i}$ has to satisfy the relation 
$p _{i}F ^{i} = 0$. 

Both the collision integral $C$ and the force $F ^{i}$ describe interactions within the many-particle system. 
While $C$ conventionally accounts for elastic binary collisions, 
we intend $F ^{i}$ to model different kinds of interactions in a simple manner. 
Strictly speaking, $F ^{i}$ should be calculated from the microscopic particle dynamics and, consequently, depend on the entire set of particle coordinates and momenta characterizing the $N$-particle system. 
Since such kind of approach appears too complicated, in fact it already  requires a solution of the many-particle dynamics, we follow here a different strategy. 
We {\it assume} $F ^{i}$ to be an effective one-particle quantity which instead of depending on the coordinates and momenta of the remaining particles, 
is supposed to depend on macroscopic fluid quantities characterizing the $N$-particle system as a whole. 
At the moment we do not specify this force. It will be determined below by general equilibrium conditions. 
Although lacking a detailed microscopic justification, the selfinteracting force concept will result in a transparent picture of the dynamics 
of the gas both macroscopically and microscopically.  

The particle number flow 4-vector 
$N^{i}$ and the energy momentum tensor $T^{ik}$ are
defined in a standard way (see, e.g., \cite{Ehl}) as 
\begin{equation}
N^{i} = \int \mbox{d}Pp^{i}f\left(x,p\right) \mbox{ , } 
\ \ \ 
T^{ik} = \int \mbox{d}P p^{i}p^{k}f\left(x,p\right) \mbox{ .} 
\label{3}
\end{equation}
The integrals in (\ref{3}) and throughout 
the paper
are integrals over the entire mass shell 
$p^{i}p_{i} = - m^{2}$. 
The entropy flow vector $S^{a}$ is given by \cite{Ehl}, \cite{IS} 
\begin{equation}
S^{a} = - \int p^{a}\left[
f\ln f - f\right]\mbox{d}P \mbox{ , }
\label{4}
\end{equation}
where we have restricted ourselves to the case of 
classical Maxwell-Boltzmann particles. 
 
Using well-known general relations (see, e.g., \cite{Stew})  
we find 
\begin{equation}
N^{a}_{;a} = \int \left(C\left[f\right] 
- m F ^{i}\frac{\partial{f}}{\partial{p ^{i}}}\right) \mbox{d}P 
\mbox{ , } \ \ 
T^{ak}_{\ ;k} =  \int p^{a}\left(C\left[f\right] 
- m F ^{i}\frac{\partial{f}}{\partial{p ^{i}}}\right) 
\mbox{d}P
\mbox{ , } 
\label{5}
\end{equation}
and 
\begin{equation}
S^{a}_{;a} = - \int \ln f 
\left(C\left[f\right] 
- m F ^{i}\frac{\partial{f}}{\partial{p ^{i}}}\right) \mbox{d}P
\mbox{ .} \label{6}
\end{equation}
In collisional equilibrium, which we
shall assume from
now on, $\ln f$ in
(\ref{6}) 
is a linear combination of the collision invariants 
$1$ and $p^{a}$.  
The corresponding equilibrium distribution function 
becomes (see, e.g., \cite{Ehl}) 
\begin{equation}
f^{0}\left(x, p\right) = 
\exp{\left[\alpha + \beta_{a}p^{a}\right] } 
\mbox{ , }\label{7}
\end{equation}
where $\alpha = \alpha\left(x\right)$ and 
$\beta_{a}\left(x \right)$ is timelike.  
Inserting the equilibrium function (\ref{7}) 
into eq.(\ref{1}) one obtains
\begin{equation}
p^{a}\alpha_{,a} +
\beta_{\left(a;b\right)}p^{a}p^{b}   
=  - m \beta _{i}F ^{i} 
\mbox{ .} \label{8}
\end{equation} 
For a vanishing force $F ^{i}$ the latter condition reduces to the  ``global '' equilibrium condition of standard relativistic kinetic theory. 
For $F ^{i} \neq 0$ condition (\ref{8}) is a ``generalized '' equilibrium condition (see \cite{WZ} and below). 

Use of Eq. (\ref{7}) in the balances (\ref{5}) yields 
\begin{equation}
N^{a}_{;a}=-m \beta _{i}\int F ^{i}f ^{0}\mbox{d}P \mbox{\ , \ \  } 
T^{ak}_{\ ;k}=-m \beta _{i}\int p^{a}F ^{i}f ^{0}\mbox{d}P \ .
\label{9}
\end{equation}

For the entropy production density we find 
\begin{equation}
S^{a}_{;a} = m \beta _{i} \int \left[\alpha + \beta _{a}p ^{a} \right]
F ^{i}
\ln f^{0} \mbox{d}P 
=  - \alpha N^{a}_{;a} 
- \beta_{a}T^{ab}_{\ ;b}
\mbox{ . }
\label{10}
\end{equation}  

With $f$ replaced by $f^{0}$ in 
(\ref{3}) and (\ref{4}), $N^{a}$, $T^{ab}$ and $S^{a}$ may be 
split with respect to the unique 4-velocity $u^{a}$ according to 
\begin{equation}
N^{a} = nu^{a} \mbox{ , \ \ }
T^{ab} = \rho u^{a}u^{b} + p h^{ab} \mbox{ , \ \ }
S^{a} = nsu^{a} \mbox{  , }
\label{11}
\end{equation}
where $h ^{ab}$ is the 
spatial projection tensor $h^{ab} = g^{ab} + u^{a}u^{b}$, 
$n$ is the particle number density, $\rho$ is the energy
density, $p$ is the equilibrium pressure and 
$s$ is the entropy per particle. 
The exact integral expressions for $n$, $\rho$, $p$ and $s$ are given
by the formulae (177) - (180) in \cite{Ehl}. 

Using (\ref{11}) and defining
\begin{equation}
\Gamma  \equiv -\frac{m}{n} \beta _{i}\int F ^{i}f ^{0} \mbox{d}P 
\mbox{ , }
\label{12}
\end{equation}
the first eq.(\ref{9}) becomes 
\begin{equation}
\dot{n} + \Theta n = n\Gamma  \mbox{ ,}\label{13}
\end{equation}
where $\Theta \equiv  u ^{a}_{;a}$ is the fluid expansion. 
It is obvious that $\Gamma$ is the particle production rate. 
Similarly, 
with the decomposition (\ref{11}) and the abbreviation
\begin{equation}
t^{a} \equiv m \beta _{i} \int p^{a}F ^{i}f ^{0} \mbox{d}P \mbox{ , }
\label{14}
\end{equation}
we obtain 
\begin{equation}
T ^{ab}_{\ ;b} + t ^{a} = 0 \ ,
\label{15}
\end{equation}
implying
\begin{equation}
\dot{\rho } + \Theta \left(\rho + p \right) = u _{a}t ^{a}\ ,
\ \ \ \ \ 
\left(\rho + p \right)\dot{u}_{a} + \nabla  _{a}p 
= - h _{ai}t ^{i}\ .
\label{16}
\end{equation}
According to (\ref{13}) and (\ref{16}) neither the particle number nor the energy-momentum are preserved in the presence of a nonvanishing $F ^{i}$. 
A varying particle number may be thought of as a result of inelastic interactions inside the medium \cite{ZMNRAS}. 
Another possibility is particle production out of the gravitational field 
\cite{Zel,Mur,Hu}. 
Source terms in the balances (\ref{16}) are equivalent to the fact that the energy-momentum tensor $T ^{ab}$ in (\ref{11}) is not conserved. 
Consequently, $T ^{ab}$ is {\it not} the energy-momentum tensor for the selfinteracting gas a a whole (see below). 

\section{Quasilinear forces and generalized equilibrium}

The four-momenta $p ^{a}$ may be decomposed into 
\begin{equation}
p ^{a} = E u ^{a} + \lambda e ^{a} \ ,
\label{17}
\end{equation}
where $e ^{a}$ is a unit spatial vector, i.e., $e ^{a}e _{a} = 1$, 
$e ^{a}u _{a} = 0$. 
Consequently, one has $E = - u _{a}p ^{a}$ and $\lambda = e _{a}p ^{a}$ 
and the 
mass shell condition $p ^{a}p _{a} = - m ^{2}$ is equivalent to 
$\lambda ^{2} = E ^{2} - m ^{2}$. 
Moreover, $h _{ab}p ^{a}p ^{b} = \lambda ^{2}$ is valid.
For the force $F ^{m}$ we write analogously 
$F ^{m} = F \left(x,p \right)u ^{m} 
+ K \left(x,p \right)e ^{m}$,  
where $F \equiv  - u _{m}F ^{m}$ and $K \equiv  e _{m}F ^{m}$. 
The requirement $p _{m}F ^{m} = 0$ will be automatically fulfilled for 
\begin{equation}
F ^{m} = \left[u ^{m} + \frac{E}{\lambda }e ^{m} \right]
F \left(x,p \right)\ .
\label{18}
\end{equation}
With the familiar identification $\beta _{m} = u _{m}/T$ it is obvious that only the part $u _{m}F ^{m} \equiv  - F$ contributes in 
the ``source terms'' (\ref{12}) and 
(\ref{14}). 
Up to now the function $F\left(x,p \right)$ is completely general. 
Specific assumptions on the functional structure of $F$ are necessary to evaluate explicitly the integrals in (\ref{12}) and (\ref{14}). 
We will {\it assume} here that the macroscopic fluid dynamics in the presence of a nonvanishing force $F ^{i}$ continues to imply only moments of the distribution function not higher than the second ones. 
It is obvious from the expressions (\ref{12}) and (\ref{14}), which act as source terms in the balances (\ref{13}) and (\ref{16}), respectively, that this requirement restricts $F$ to depend on $p ^{a}$ only linearly. 
The most general linear expression for $F$ is 
\begin{equation}
F \left(x,p \right) = F _{0}\left(x \right) + F _{E}\left(x \right)E 
+ F _{\lambda }\left(x \right)\lambda \ .
\label{19}
\end{equation}
Because of the additional dependence on $p ^{m}$ through 
$E = - u _{m}p ^{m}$ and $\lambda = e _{m}p ^{m}$ in the bracket in front of 
$F \left(x,p \right)$ in (\ref{18}), we call the force 
(\ref{18}) with 
the specific dependence (\ref{19}) ``quasilinear''.  
With this structure of the force 
the equilibrium condition (\ref{8})  becomes 
\begin{equation}
p ^{a}\alpha _{,a} + \beta _{\left(a;b \right)}p ^{a}p ^{b} = 
\frac{m}{T}
\left[ F _{0}\left(x \right) + F _{E}\left(x \right)
\left(-u _{a}p ^{a} \right) 
+ F _{\lambda }\left(x \right)\left(e _{a}p ^{a} \right)  \right]\ .
\label{20}
\end{equation}
Introducing the decomposition (\ref{17}) also on the left-hand side of the last equation we find for the parameters $\alpha$ and $\beta $ of the equilibrium distribution function
\begin{equation}
\dot{\alpha}=\frac{m}{T}F _{E}\ ,\ \ \ \ 
e ^{a}\nabla  _{a}\alpha = \frac{m}{T}F _{\lambda }\ ,
\label{21}
\end{equation}
where $\nabla  _{a}\alpha \equiv  h _{a}^{b}\alpha _{,b}$, and
\begin{equation}
\beta_{(a;b)}= \phi (x)g_{ab}\ , 
\label{22}
\end{equation}
with $\phi  = - F _{0}/(m T)$. 
The parameter $\alpha$ changes both in space and time while it is constant 
in the standard ``global'' equilibrium corresponding to the force-free case. 
With $\beta _{a}=u _{a}/T$ 
the conformal Killing vector 
(CKV) property (\ref{22})  implies the relation
\begin{equation}
\phi = \frac{1}{3}\frac{\Theta }{T}\ ,
\label{23}
\end{equation}
which fixes the function $F _{0}$:
\begin{equation}
F _{0} = - \frac{m}{3}\Theta \ .
\label{24}
\end{equation}
$F _{0}$ vanishes for radiation ($m = 0$). 

It is remarkable that through $F _{0}$ the force depends on the fluid expansion, a quantity characterizing the gas as a whole on the macroscopic level. Similar dependences on macroscopic fluid quantities will be found below for $F _{E}$ and $F _{\lambda }$. 
{\it Since both the microscopic particle four-momentum $p ^{a}$ and macroscopic fluid quantities, characterizing the system as a whole, 
enter the four-force, the latter 
represents a selfinteraction of the gas.}  

The ``sources'' (\ref{12}) and (\ref{14}) in the balances 
(\ref{13}) and (\ref{16}), respectively, may be expressed in terms of  
$F _{0}$, $F _{E}$ and $F _{\lambda }$.  
For $\Gamma $ defined by (\ref{12}) we find 
\begin{equation}
\Gamma = \frac{m}{T}\left(\frac{M}{n}F _{0} + F _{E} \right)\ ,
\label{25}
\end{equation}
where  $M$ is the zeroth moment of the distribution function,  
$M \equiv \int \mbox{d}Pf ^{0}\left(x, p\right) 
= \left(\rho - 3p \right)/m ^{2}$. 
Analogously, $t ^{a}$ in (\ref{14}) may be written as 
\begin{equation}
t ^{a} = - \frac{m}{T}\left[F _{0}N ^{a} 
- F _{E}u _{n}T ^{an} 
+ F _{\lambda }e _{n}T ^{an} \right]\ ,
\label{26}
\end{equation}
implying
\begin{equation}
u _{a}t ^{a} = \frac{m}{T}\left(nF _{0} + \rho F _{E} \right)\ ,
\ \ \ \ \ 
h _{ca}t ^{a} = - \frac{m}{T}p F _{\lambda }e _{c}\ .
\label{27}
\end{equation}

The Gibbs equation 
\begin{equation}
T \mbox{d}s = \mbox{d} \frac{\rho }{n} + p \mbox{d}\frac{1}{n}\ ,
\label{28}
\end{equation}
provides us with 
\begin{equation}
nT \dot{s} = \dot{\rho } - \left(\rho + p \right)\frac{\dot{n}}{n} 
= u _{a}t ^{a} - \left(\rho  + p \right)\Gamma 
\label{29}
\end{equation}
for the time change $\dot{s}$ of the entropy per particle. 
Here we have used 
the balances (\ref{13}) and (\ref{16}) 
to obtain the second equation (\ref{29}) . 

With the identification $\alpha = \mu /T$ where $\mu $ is the chemical potential 
and \cite{Groot}
\begin{equation}
s = \frac{\rho + p}{nT} - \frac{\mu }{T}
\label{30}
\end{equation}
the entropy production (\ref{10}) becomes
\begin{equation}
S ^{a}_{;a} = n \Gamma s + n \dot{s}\ .
\label{31}
\end{equation}

The equations of state for a gas may be generally written as (see, e.g., \cite{Groot})
\begin{equation}
p = nT \ ,\ \ \ \ \ \ \ \ \ 
\rho = \rho \left(n,T \right)\ .
\label{32}
\end{equation}
Differentiating the latter relation and using the balances (\ref{13}) and (\ref{16}) with (\ref{29}) we obtain
\begin{equation}
\frac{\dot{T}}{T} = - \left(\Theta - \Gamma  \right) 
\frac{\partial p}{\partial \rho } 
+ \frac{n \dot{s}}{\partial \rho / \partial T}\ ,
\label{33}
\end{equation}
where the abbreviations
\[
\frac{\partial{p}}{\partial{\rho }} \equiv  
\frac{\left(\partial p/ \partial T \right)_{n}}
{\left(\partial \rho / \partial T \right)_{n}} \ ,
\ \ \ \ \ \ \ \ 
\frac{\partial{\rho }}{\partial{T}} \equiv  
\left(\frac{\partial \rho }{\partial T} \right)_{n}
\]
have been used. 

Restricting ourselves to the case $\dot{s} = 0$, relation (\ref{31}) implies that there is entropy production only due to the enlargement of the phase space (cf \cite{TZP,ZTP,WZ}). 
States of the system in which the gas particles (i) are governed by the 
equilibrium distribution function $f ^{0}$, (ii) the parameters $\alpha$ and 
$\beta _{a}$ in $f ^{0}$ obey (\ref{20}), equivalent to (\ref{21}) and (\ref{22}), 
and (iii) the entropy production $S ^{a}_{;a}$ is given by 
$S ^{a}_{;a} = ns \Gamma$ with $\Gamma \geq 0$, are 
called ``generalized equilibrium'' states \cite{WZ}. 
We recall that the restriction $\dot{s} = 0$ refers to processes involving the force $F ^{i}$. Conventional dissipative processes due to deviations from collisional equilibrium have already  been excluded by requiring the distribution function to have the  structure (\ref{7}) which makes the contribution of $C \left[f \right]$ in the entropy production (\ref{6}) vanish. 

The condition $\dot{s} = 0$ relates the originally independent terms 
$u _{a}t ^{a}$ and $\Gamma $:
\begin{equation}
\dot{s} = 0 \quad\Rightarrow\quad 
u _{a}t ^{a} = \left(\rho + p \right)\Gamma \ .
\label{34}
\end{equation}
Moreover, the CKV property (\ref{22}) of $\beta _{i} = u _{i}/T$ implies 
\begin{equation}
\frac{\dot{T}}{T} = - \frac{\Theta }{3}\ .
\label{35}
\end{equation}
The last relation is consistent with 
(\ref{33}) for $\dot{s} = 0$ if 
\begin{equation}
\Gamma = \left(1 - \frac{1}{3}
\frac{\partial{\rho }}{\partial{p}} \right)\Theta \ ,
\label{36}
\end{equation}
where \cite{TZP}
\begin{equation}
\frac{\partial{\rho }}{\partial{p}} = \left(\frac{m}{T} \right)^{2} 
- 1 + 5 \frac{\rho + p}{n T} 
- \left(\frac{\rho + p}{n T} \right)^{2}\ .
\label{37}
\end{equation}
Since $\frac{\partial{\rho }}{\partial{p}}\leq 3$ we have 
$\Gamma \geq 0$ in an expanding universe, which is equivalent to $S ^{a}_{;a} \geq 0$ by virtue of  (\ref{31}). 
{\it The selfinteracting force $F ^{i}$ provides us with a nonnegative expression for the entropy production in a similar sense in which 
Boltzmann's $H$ theorem guarantees $S ^{a}_{;a} \geq 0$ for elastic binary collisions}. 
Since relation (\ref{36}) is a consequence of the equilibrium conditions it implies also the statement that 
{\it except for $p=\rho /3$ (radiation) the generalized equilibrium 
conditions require a nonvanishing particle production rate $\Gamma $.} 
If there is no particle production or if the rate at which the particle number changes is different from (\ref{36}), an equilibrium such as discussed here is impossible. 

The Gibbs-Duhem equation (see, e.g., \cite{Groot})
\begin{equation}
\mbox{d} p = \left(\rho + p \right)\frac{\mbox{d} T}{T} 
+ n T \mbox{d} \left(\frac{\mu }{T} \right)
\label{38}
\end{equation}
together with the first equation of state (\ref{32}) and 
equations  (\ref{13}) and (\ref{35}) yields 
\begin{equation}
\left(\frac{\mu }{T} \right)^{\displaystyle \cdot} 
= \Gamma  - \left(1 - \frac{\rho }{3p} \right)\Theta \ .
\label{39}
\end{equation}
Inserting here the particle production rate (\ref{36}), the first 
equilibrium 
condition (\ref{21}) fixes $F _{E}$:
\begin{equation}
F _{E} = \frac{T}{3m}\left(\frac{\rho }{p} - 
\frac{\partial{\rho }}{\partial{p}} \right)\Theta 
= - \frac{T}{m}\left(\frac{\rho }{p} 
- \frac{\partial{\rho }}{\partial{p}} \right)
\frac{F _{0}}{m}\ .
\label{40}
\end{equation}
For radiation, i.e. $\rho = 3nT$, the quantity $F _{E}$ vanishes. 
For nonrelativistic matter with $\rho = nm + \frac{3}{2}nT$, 
$m \gg T$ one obtains
\begin{equation}
F _{E} = \frac{\Theta }{3}
\mbox{\ \ \ \ \ \ \ \ \ \ }
\left(m \gg T  \right) \ .
\label{41}
\end{equation}
As to the $\lambda $ part of the force $F _{\lambda }$, 
the second equilibrium condition (\ref{21}) allows us to write
\begin{equation}
F _{\lambda } = \frac{T}{m}\sqrt{\nabla  ^{a}\left(\frac{\mu }{T}\right)
\nabla  _{a}\left(\frac{\mu }{T}  \right)} \ .
\label{42}
\end{equation}
From the Gibbs-Duhem relation (\ref{38}) one finds that the spatial gradient 
$\nabla  _{a}\left(\mu /T \right)$ may be expressed by the 
fractional gradients of $n$ and $T$, 
\begin{equation}
\nabla  _{a}\left(\frac{\mu }{T} \right) = 
\frac{\nabla  _{a}n}{n} - \frac{\rho }{p}\frac{\nabla  _{a}T}{T}\ .
\label{43}
\end{equation}
Since for radiation $n \propto T ^{3}$  is valid, one has  
$\nabla  _{a}\left(\mu /T \right) = 0$, i.e.,  also the $\lambda $-part of 
the force vanishes in this limit. The motion of ultrarelativistic particles 
is force free. This corresponds to the vanishing of the source terms 
(\ref{25}) and (\ref{26}) in the macroscopic balances (\ref{13}) and 
(\ref{15}), respectively. 

For nonrelativistic matter we obtain
\begin{equation}
\nabla  _{a}\left(\frac{\mu }{T} \right) = - \frac{m}{T}
\frac{\nabla  _{a}T}{T}
\mbox{\ \ \ \ \ \ \ \ \ \ }
\left(m \gg T  \right) \ ,
\label{44}
\end{equation}
and the $\lambda $ part of the force becomes
\begin{equation}
F _{\lambda } = \sqrt{\frac{\nabla  ^{a}T \nabla  _{a}T}{T ^{2}}} 
= \sqrt{\dot{u}^{a}\dot{u}_{a}} 
\mbox{\ \ \ \ \ \ \ \ \ \ }
\left(m \gg T  \right) \ ,
\label{45}
\end{equation}
where the relation $\nabla  _{a}T/T + \dot{u}_{a} = 0$, which follows from the CKV property (\ref{22}),  
has been used. 

Combining (\ref{24}), (\ref{40}) and (\ref{42}), we find for the quantity $F \equiv  - u _{m}F ^{m}$ in (\ref{19})  
\begin{equation}
F \left(x,p \right) = - \frac{m}{3}\Theta  - \frac{T}{m} 
\left[\left(\frac{\rho }{p} 
- \frac{\partial{\rho }}{\partial{p}} \right) \frac{\Theta }{3}
u _{a} - \sqrt{\nabla  ^{a}\left(\frac{\mu }{T} \right)
\nabla  _{a}\left(\frac{\mu }{T} \right)} e _{a}\right]p ^{a}\ .
\label{46}
\end{equation}
Its first two parts are proportional to the fluid expansion while the last 
part is proportional to the spatial gradient of $\mu /T$. 
While for ultrarelativistic particles ($\rho = 3nT$) each 
of the three parts in (\ref{46}) 
vanishes separately, the force is different from zero 
for any other equation of state. 
We conclude that 
{\it a system of particles with nonzero mass $m$ is in 
generalized equilibrium 
if inbetween the elastic, binary collisions the particles move in the force field (\ref{18}) with (\ref{46})}. 
In the nonrelativistic limit $m \gg T$ the expression (\ref{46})  reduces to 
\begin{equation}
F \left(x,p \right) = \left(E - m \right)\frac{\Theta }{3} 
+ \lambda \sqrt{\dot{u}^{a}\dot{u}_{a}} 
\mbox{\ \ \ \ \ \ \ \ \ \ }
\left(m \gg T  \right) \ . 
\label{47}
\end{equation}
We recall that this force is not an external force but determined by the macroscopic fluid quantities of the 
system of colliding gas particles themselves. 
The force $F ^{m}$ represents a selfinteraction of the gas under the equilibrium conditions (\ref{21}) and (\ref{22}). 

The discussion so far allows us to reconsider earlier work 
(\cite{TZP,ZTP,WZ}) on kinetic theory and particle production based on 
a Boltzmann equation modified by a ``source'' term which was assumed 
to be linear in the distribution function 
with an additional linear dependence on the particle four-momenta 
of the factor in front of $f ^{0}$ 
(see formulae (37) and (38) in \cite{TZP}). 
While the second assumption is equivalent to our eq. (\ref{19}), 
there is no need for a conterpart of the first one in the present paper. 
Although formally similar, the description in terms of selfinteracting forces is physically more transparent than the somewhat vague concept of a ``source term'' in Boltzmann's equation. 
For comparison, we list up the correspondences between the spacetime functions $\nu$, $\tau $ and $\sigma $, characterizing the mentioned source term in \cite{TZP,ZTP,WZ}, and the quantities $F _{0}$, $F _{E}$ and $F _{\lambda }$ of the present paper:
\begin{equation}
\nu \Leftrightarrow \frac{m}{T}F _{0}\ ,
\ \ \ \ \ 
\frac{1}{\tau } \Leftrightarrow \frac{m}{T}F _{E}\ ,
\ \ \ \ \ 
\frac{1}{\sigma } \Leftrightarrow \frac{m}{T}F _{\lambda }\ .
\label{47a}
\end{equation}
The ``sources'' $\nu$, $1/\tau $ and $1/\sigma $  
(see \cite{TZP,ZTP,WZ}) 
may be understood in 
terms of selfinteracting forces on the microscopic gas particles. 

The main advantage of the force concept compared with the ``source term'' approach is the following. 
Once determined by the equilibrium properties of the gas, the force may now be used to reconsider the microscopic equations of motion. 
With the splitting (\ref{17})  the left-hand side of the second equation (\ref{2}) may be written as 
\[
\frac{D p ^{i}}{d \tau } = \frac{d E}{d \tau }u ^{i} 
+ E \frac{D u ^{i}}{d \tau } + \frac{d \lambda }{d \tau }e ^{i} 
+ \lambda \frac{D e ^{i}}{d \tau }\ .
\]
Contraction with $u _{i}$ yields 
\[
u _{i}\frac{D p ^{i}}{d \tau } = - \frac{d E}{d \tau }   
+ \lambda u _{i}\frac{D e ^{i}}{d \tau } 
= - \frac{d E}{d \tau }   
- \lambda e ^{i}\frac{D u _{i}}{d \tau }\ .
\]
Use of  
\[
\frac{D u ^{i}}{d \tau } = u ^{i}_{;n}\frac{p ^{n}}{m}
\]
allows us to write  
\[
u _{i}\frac{D p ^{i}}{d \tau } = - \frac{d E}{d \tau }   
- \frac{\lambda E}{m}e ^{i}\dot{u}_{i} 
- \frac{\lambda ^{2}}{m}e ^{i}e ^{n}u _{i;n}\ .
\]
Applying here the decomposition \cite{Ehlers,Ellis}
\begin{equation}
u _{i;n} = - \dot{u}_{i}u _{n} + \sigma _{in} +  \omega _{in}  
+ \frac{\Theta }{3}h _{in}\ ,
\label{47b}
\end{equation}
where
$\omega_{ab} = h_{a}^{c}h_{b}^{d}u_{\left[c;d\right]}$ and 
$\sigma _{ab} = h_{a}^{c}h_{b}^{d}u_{\left(c;d\right)} 
- \frac{\Theta }{3}h _{ab}$,  
the equation of motion (\ref{2}), projected in direction of $u ^{i}$, 
\begin{equation}
u _{i}\frac{D p ^{i}}{d \tau } = u _{i}F ^{i} = - F, 
\label{47c}
\end{equation}
may generally be written as 
\begin{equation}
\frac{d E}{d \tau } + \frac{\lambda E}{m}e ^{i}\dot{u}_{i} 
+ \frac{\lambda ^{2}}{m}e ^{i}e ^{n}\sigma _{in} 
+ \frac{\lambda ^{2}}{3m}\Theta  = F \ .
\label{47d}
\end{equation}
In the following section we will integrate this equation for specific cases.

\section{Fluid and particle dynamics} 
Obviously, all relations of the preceeding section are applicable on an arbitrary, given gravitational background. 
Considering the gravitational field of such kind of selfinteracting equilibrium system requires further investigations. 
Since the energy momentum tensor $T ^{ab}$ in (\ref{3}) and (\ref{11}) is not conserved, it is not a suitable quantity on the right-hand side of Einstein's field equations.  
The question arises whether one may introduce an effectively conserved energy-momentum tensor $\tilde{T}^{ak}$ satisfying 
$\tilde{T}^{ak}_{\ ;k} \equiv  T ^{ak}_{\ ;k} + t ^{a} = 0$, where the ``source'' $t ^{a}$ is mapped onto suitable components of the energy-momentum tensor $\tilde{T}^{ak}$. 
This probem has been solved previously \cite{TZP,ZTP,WZ} and the corresponding results may be applied immediately. 
The essential point is that the balances (\ref{16}) following from 
(\ref{15}) are identical to the balances 
\begin{equation}
\dot{\rho } + \Theta \left(\rho + p + \pi  \right) = 0 \ ,
\ \ \ \ \ \ 
\left(\rho + p + \pi  \right)\dot{u}_{a} 
+ \nabla  _{a}\left(p + \pi  \right) = 0
\label{48}
\end{equation}
following from $\tilde{T}^{ab}_{\ ;b} = 0$ with 
\begin{equation}
\tilde{T}^{ab} = \rho u ^{a}u ^{b} + \left(p + \pi  \right)h ^{ab}
\label{49}
\end{equation}
and the identifications
\begin{equation}
u _{a}t ^{a} = - \Theta \pi \ ,
\ \ \ \ \ 
h _{ai}t ^{i} = \pi \dot{u}_{a} + \nabla  _{a}\pi \ .
\label{50}
\end{equation}
The source terms $t^{a}$ in (\ref{15}) may  consistently be 
mapped onto an effective viscous pressure $\pi $ of a locally conserved 
energy-momentum tensor $\tilde{T}^{ab}$. 
This viscous pressure is determined by eqs. (\ref{34}) and (\ref{36}): 
\begin{equation}
\pi = - \left(\rho + p \right)\left(1 - \frac{1}{3}
\frac{\partial{\rho }}{\partial{p}} \right). 
\label{51}
\end{equation}
The quantity $\tilde{T}^{ab}$ in (\ref{49}) with (\ref{51}) may be regarded as the effective energy-momentum tensor of the selfinteracting many-particle system. 

The first purpose of this section is to demonstrate how $\pi $ is determined by the selfinteracting force.
Combining (\ref{27}) with (\ref{34}) and (\ref{36}) yields
\begin{equation}
\pi = - \frac{m}{T}\left(n F _{0} + \rho F _{E} \right)\Theta ^{-1}\ .
\label{52}
\end{equation} 
Applying (\ref{34}) with 
(\ref{25}) and (\ref{40}) as well as (\ref{24}),  
the following equivalent representations for $\pi $ are possible:  
\begin{equation}
\pi = \frac{m}{T}\left(\rho + p \right)
\left[\frac{M}{n}F _{0} + F _{E} \right]\Theta ^{-1} 
= - \frac{1}{3}\left(\frac{m}{T} \right)^{2}
\left[\frac{\rho - 3p}{nm}\frac{\rho }{nm} - 1\right]\ .
\label{53}
\end{equation} 
While (\ref{36}) and (\ref{51}) are previous results \cite{TZP,ZTP,WZ},  
the expressions (\ref{25}), (\ref{52}) and (\ref{53}) are new. 
The latter 
are equivalent to (\ref{36}) and (\ref{51}) but differently from these previously obtained formulae the relations 
(\ref{25}) and (\ref{52}) (or (\ref{53})) establish a direct connection  
between the  force on the gas particles and the particle production 
rate $\Gamma $ and the effective viscous pressure $\pi $, respectively. 
The deviations from the standard perfect fluid behaviour, represented by 
$\pi $, are immediately connected to a microscopic 
selfinteraction of the gas, 
which governs the particle motion according to (\ref{47d}). 
 
The backreaction of a nonvanishing $\pi $ on the fluid dynamics (\ref{48})
depends on the equations of state. 
There is no backreaction for $\rho = 3nT$. But for any other equation of state the force $F ^{m}$ and, consequently, the particle production rate $\Gamma $ and the effective viscous pressure $\pi $, are different from zero. 
The backreaction is largest for $m \gg T$, i.e., for nonrelativistic matter. 
Introducing a length scale $a$ according to 
$\Theta \equiv  3 \dot{a}/a$ one obtains 
\begin{equation}
\Gamma = \frac{\Theta }{2}\ ,
\mbox{\ \ \ \ }
\pi = - \frac{\rho }{2}\ ,
\mbox{\ \ \ \ }
n \propto a ^{-3/2}\ , 
\mbox{\ \ \ \ }
\rho \propto a ^{-3/2} 
\mbox{\ \ \ \ } 
\left(m \gg T \right) \ .
\label{55}
\end{equation}
The temperature changes according to (\ref{35}) for any equation of state. 
The familiar behaviour of nonrelativistic matter characterized by 
$n \propto a ^{-3}$, $T \propto a ^{-2}$ and 
$\rho \propto a ^{-3}$ for $\Gamma = \pi = 0$ is modified considerably. 
All thermodynamic quantities decrease more slowly for $\pi \neq 0$ since the decay of 
$n$, $T$ and $\rho $ due to the expansion is counteracted by corresponding production terms reflecting the selfinteraction of the gas particles on a macroscopic level. 
Assuming that matter in this kind of equilibrium dominates the cosmological dynamics, the expansion of the universe is modified as well \cite{ZTP,WZ}. 
In a homogeneous and isotropic universe the lenght scale $a$ coincides with 
the scale factor of the Robertson-Walker metric and,   
in the spatially flat case, obeys 
the equation 
\begin{equation}
3 \frac{\dot{a}^{2}}{a ^{2}} = \kappa \rho \ ,
\label{56}
\end{equation}
where $\kappa$ is Einstein's gravitational constant. 
For radiation with $\pi = 0$ we recover, of course, $a \propto t ^{1/2}$. 
Inserting, however, the energy density 
$\rho $ from Eq. (\ref{55}) into Eq. (\ref{56}) we find that 
the scale factor behaves such as
\begin{equation}
a \propto t ^{4/3} 
\mbox{\ \ \ \ \ \ \ \ \ \ \ \ \ \ }
\left(m \gg T \right) \ ,
\label{57}
\end{equation}
i.e., $\ddot{a} > 0$ instead of the familiar $a \propto t ^{2/3}$ with  
$\ddot{a} < 0$ for $\rho \propto a ^{-3}$ corresponding to $\Gamma = 0$. 
{\it Generalized equilibrium of a selfinteracting 
massive gas universe implies power law inflation.} 
The possibility of an accelerated expansion of the 
universe may be traced back to specific forces on the microscopic constituents of the cosmic medium.  

Since these forces are explicitly known, we may solve the microscopic particle dynamics as well. 
For homogeneous and isotropic universes, characterized by 
$\dot{u}_{i} = \sigma _{in} = 0$, equation (\ref{47d}) reduces to 
\begin{equation}
\frac{d E}{d \tau } + \frac{\lambda ^{2}}{3m}\Theta  = F \ .
\label{68}
\end{equation}
With $d \tau = d t \left(m/E \right)$, $\lambda ^{2} = E ^{2} - m ^{2}$, 
and $d E/d t \equiv  \dot{E}$, the last equation is equivalent to 
\begin{equation}
\frac{\left(E ^{2} - m ^{2} \right)^{\displaystyle \cdot}}
{E ^{2} - m ^{2}} + \frac{\left(a ^{2} \right)^{\displaystyle \cdot}}
{a ^{2}} = \frac{2m}{E ^{2} - m ^{2}}F \ .
\label{69}
\end{equation}
In the limit $F = 0$ (geodesic motion) we obtain 
\begin{equation}
E ^{2} - m ^{2} = \lambda ^{2} \propto a ^{-2}\ ,
\mbox{\ \ \ }
\left(F = 0 \right)\ ,
\label{70}
\end{equation}
implying the expected behaviour $E \propto a ^{-1}$ for massless particles (photons) while the nonrelativistic energy $\epsilon \equiv  E - m$ with $\epsilon \ll m$ of massive particles decays as $\epsilon \propto a ^{-2}$. 
The most interesting case (\ref{47}) simplifies to 
$F = \left(E - m \right)\Theta /3$ for $\dot{u}_{i} = 0$. Under these conditions which macroscopically are characterized by 
(\ref{55}) and (\ref{57}), 
the solution of equation (\ref{68}) is 
\begin{equation}
E - m  \propto  a ^{-1} \ ,
\mbox{\ \ \ \ \ }
\left(F = \left(E - m \right)\frac{\Theta }{3} \right)\ . 
\label{70a}
\end{equation}
The nonrelativistic energy of massive particles under generalized equilibrium conditions in a (quasi-)linear force field decays linearly with the cosmic scale factor, i.e., the selfinteracting force makes nonrelativistic particles behave like radiation. 
{\it Massive particles under the action of a force field (\ref{18}) with 
$F = \left(E - m \right)\Theta /3$ behave like photons}. 
This is the microscopic counterpart of the statement that radiation and nonrelativistic matter may be in equilibrium in the expanding universe, provided the number of matter particles increases at a specific rate 
\cite{ZTP}. 
The solution (\ref{70a}) for the microscopic particle dynamics together with the corresponding macroscopic fluid behaviour (\ref{55}) and (\ref{57}) constitutes a nontrivial, exactly solvable model of a gas universe in generalized equilibrium.

\section{Summary}

We have investigated the kinetic theory of a classical relativistic gas of  particles moving  
under the influence of a simple, selfinteracting force inbetween 
equilibrium establishing 
elastic, binary collisions.  
The concept of a selfinteracting force was shown to provide a comprehensive picture of the gas dynamics both macroscopically and microscopically. 
On the macroscopic level it gives rise to an effective viscous pressure which, on the microscopic level, corresponds to a deviation from geodesic particle motion. 
A deeper understanding of previously obtained results on the backreaction of particle production processes on the cosmological dynamics was obtained. 
We established an exactly solvable model according to which massive gas particles under the action of a specific force behave like photons. 
A universe of gas particles of this kind is necessarily in a state of accelerated expansion corresponding to a behaviour 
$a \propto t ^{4/3}$ of the scale factor $a$ of the Robertson-Walker metric. 
\\
\ \\
\ \\
{\bf Acknowledgements}\\
\ \\
This paper was supported by the 
Deutsche Forschungsgemeinschaft.  \\
\ \\

\end{document}